\documentclass[a4paper,11pt]{article}
\usepackage{pos}
\usepackage{macros}
\usepackage{slashed}

\title{Cutoff effects of the gradient flow for fermions}
\ShortTitle{Cutoff effects of gradient flow}

\author*[a]{Andrea Shindler}

\affiliation[a]{Facility for Rare Isotope Beams \& Physics Department, Michigan State University, \\ 
East Lansing, Michigan 48824, USA}

\emailAdd{shindler@frib.msu.edu}

\abstract{I analyze cutoff effects of the gradient flow for Wilson-type fermions.
I show that with a proper choice of the higher dimensional fields in the 
Symanzik effective theory, O($a$) improvement of the action is achieved 
changing the initial conditions of the gradient flow equation.}

\FullConference{%
  The 39th International Symposium on Lattice Field Theory (Lattice2022),\\
  8-13 August, 2022 \\
  Bonn, Germany 
}

\begin{document}
\maketitle

\section{Introduction}

In this contribution I perform an analysis of O($a$) cutoff effects
of the gradient flow for Wilson-type fermions~\cite{Luscher:2013cpa}.
For a recent review on the applications to renormalization of the gradient flow 
see Ref.~\cite{Shindler:2022tlx}.
Discretization effects of the gradient flow for gauge 
fields~\cite{Narayanan:2006rf,Luscher:2010iy,Luscher:2011bx} 
have been studied for example in Refs.~\cite{Fodor:2014cpa,Ramos:2015baa}.
O($a$) cutoff effects affecting correlation functions containing flowed 
fermion fields have been analyzed in Ref.~\cite{Luscher:2013cpa},
where special improvement terms, 
needed to improve correlation functions of flowed fermion fields, 
have been derived.
In the case of flowed correlation functions 
the Symanzik effective theory, beside the usual clover improvement term 
proportional to $\csw$, contains an additional term proportional to $\cfl$.
In this proceedings I discuss the reason for the presence of such additional term.
I also show that with a proper choice of the higher dimensional 
fields the theory can be alternatively improved modifying the initial conditions of the 
gradient flow equations.

\section{Cutoff effects of the gradient flow for fermions}
\label{sec:gf}
The evolution with the flow time $t$ for fermions is given by~\cite{Luscher:2013cpa}
\bea 
&& \partial_t \chi(x,t) = \Delta \chi(x,t)\,, \qquad 
\partial_t \bar\chi(x,t) = \bar\chi(x,t) \overleftarrow{\Delta}\,, \label{eq:gf} \\
&& \chi(x,t=0) = \psi(x)\,, \qquad \chibar(x,t=0) = \psibar(x)\,, \nonumber
\eea 
where $\Delta = D_\mu D_\mu$ and the covariant derivative $D_\mu = \partial_\mu + B_\mu$
contain the flowed gauge field $B_\mu(t)$.
The dynamics of correlation functions containing flowed fermion fields, $\chi$ and
$\chibar$, can be described introducing an extra-dimension to the theory, 
for the flow time $t$, and introducing suitable Lagrange multipliers,
that, once integrated out, constrain the flowed fields to satisfy the 
appropriate flow equations.
The action of the $4+1$ dimensional theory reads
\be 
S = S_{\text{G}} + S_{\text{G,fl}} + S_{\text{F}} + S_{\text{F,fl}}\,,
\ee
where $S_{\text{G}}+ S_{\text{F}}$ is the standard QCD action and 
$S_{\text{G,fl}}$ contains the Lagrange multipliers for the gauge fields 
discussed for example in Refs.~\cite{Luscher:2010iy,Ramos:2015baa}.
For the fermion fields one has
\be 
S_{\text{F,fl}} = \int_0^\infty dt~\int d^4x~
\left[\lambdabar(x,t)(\partial_t - \Delta)\chi(x,t)+
\chibar(x,t) \left(\overleftarrow{\partial_t} - \overleftarrow{\Delta}\right)
\lambda(x,t)\right]\,, 
\ee 
where $\lambda$ and $\lambdabar$ are the Lagrange multipliers that, 
once integrated out, impose to the flowed fermion fields to satisfy Eqs.~\eqref{eq:gf}.
The energy-dimension of $\lambda$ and $\lambdabar$ is $5/2$.
With the local formulation it is possible to demonstrate the
renormalizability of the modified 
theory~\cite{Luscher:2010iy,Luscher:2011bx,Luscher:2013cpa} 
and discuss chiral symmetry and related Ward 
identities~\cite{Luscher:2013cpa,Shindler:2013bia,Bar:2013ora}.

The discretization of the gauge action, provided it preserves the standard
symmetries, is not relevant for this discussion.
I choose a Wilson-type\footnote{With Wilson-type discretization I denote all lattice actions 
based on the Wilson action, such as clover fermions.} discretization for the 
fermion part of the action
and the flow time part of the fermion action is discretized with a 
step $\epsilon$ ($t=n \epsilon$) 
\be 
S_{\rm{F,fl}} = \epsilon\sum_{n\ge 0} a^4 \sum_x \left[\lambdabar(x,t)\left(\partial_t - \nabla^2 \right)
\chi(x,t) + \chibar(x,t)\left(\overleftarrow{\partial}_t - 
\overleftarrow{\nabla}^2\right)\lambda(x,t)\right]\,,
\label{eq:SFfl_latt}
\ee 
where $\nabla^2=\nabla_\mu^* \nabla_\mu$ with $\nabla_\mu$ ( and $\nabla_\mu^*$) 
the flowed forward (and backward) lattice covariant derivatives.
The discrete derivative with respect with the flow time is given by 
\be 
\partial_t\chi(x,t) = \frac{1}{\epsilon}\left(\chi(x,t+\epsilon) - \chi(x,t)\right)\,.
\ee 

To analyze cutoff effects it is convenient to describe the 
theory close to the continuum limit with an effective continuum theory, 
the so-called Symanzik effective theory, 
with higher dimensional fields multiplying powers of the lattice
spacing~\cite{Symanzik:1983dc,Luscher:1996sc}. 
The classification of the higher dimensional fields is obtained 
using standard discrete and chiral symmetry transformation 
properties of the fermion fields and the Lagrange multipliers~\cite{Shindler:2013bia}.

An analysis of the Symanzik effective theory for fermions has already been performed
in Ref.~\cite{Luscher:2013cpa}, and it is given by 
\be 
S_{\rm eff}[B,\chi,\chibar] = S_0[B,\chi,\chibar]+ a S_{1} + O(a^2)\,,
\label{eq:eff_action}
\ee
where $S_0$ denotes the target continuum theory
with renormalized parameters, and $S_1$ contains 
higher dimensional fields.

The O($a$) cutoff effects in the lattice action 
are distinguished in $S_{1,\b}$ arising from the $t=0$ boundary,
and $S_{1,\fl}$, arising from the bulk of the $4+1$ dimensional 
theory
\be 
S_{1,\b} = \int d^4x~\sum_{i=1}^{n_\b} O_i(x)\,,
\qquad 
S_{1,\fl} = \int_0^\infty dt \int d^4x~\sum_{i=1}^{n_\fl} Q_i(x,t)\,.
\ee
The fields $Q_i(t,x)$ and $O_i(x)$ are made of space-time and/or flow-time
derivatives and the fundamental 
degrees of freedom of the theory, including the Lagrange multipliers.
To keep the action with zero dimension 
the fields $Q_i(t,x)$ must have dimension $7$ while $O_i(x)$ dimension $5$.

It is sufficient to improve classically the bulk action thanks 
to the observation that in perturbation theory flowed correlation functions
generate only ``tree diagrams''~\cite{Luscher:2011bx}.
The standard Symanzik improvement program 
can be applied to the boundary term $S_{1,\b}$.

A classical expansion in 
powers of $a$ of the lattice fermion action~\eqref{eq:SFfl_latt}
dictates the form of $S_{1,\fl}$.
Expanding the covariant laplacian $\nabla^* \nabla$ 
\be 
\nabla^*\nabla  = D_\mu D_\mu\left(1+\frac{a^2}{12}D_\mu D_\mu\right) + O(a^3)\,,
\label{eq:nabla_lat}
\ee 
one obtains the expected result that 
the leading corrections to the bulk action are of O($a^2$), i.e. 
$S_{1,\fl}=0$ and the first non-leading term of the effective theory 
is $S_{2,\fl}$.
Modifying the covariant derivatives following Eq.~\eqref{eq:nabla_lat}, 
$\nabla_\mu^*\nabla_\mu \rightarrow 
\nabla_\mu^*\nabla_\mu\left(1-\frac{a^2}{12}\nabla_\mu^*\nabla_\mu\right)$,
removes the O($a^2$) stemming from the gradient flow equation~\cite{Battelli:2022kbe}.
In this work I am only considering O($a$) cutoff effects, but
it could become useful to monitor the continuum limit to include or
exclude the O($a^2$) corrections to the flow equation.
For this reason I define later a different gradient flow equation
that includes the extra term in Eq.~\eqref{eq:nabla_lat}. 

For a single flavor
the $D=5$ fields contributing to the boundary term
$S_{1,\b}$ are
\bea 
O_1(x) &=& \psibar(x)\sigma_{\mu\nu}G_{\mu\nu}(x)\psi(x)\,, \\
O_2(x) &=& \psibar(x)D_\mu D_\mu\psi(x) + 
\psibar(x)\overleftarrow{D_\mu}\overleftarrow{D_\mu}\psi(x)\,, \\
O_3(x) &=& m\Tr\left[G_{\mu\nu}G_{\mu\nu}\right]\,, \\
O_4(x) &=& m \psibar(x)\left[\gamma_\mu D_\mu - 
\gamma_\mu\overleftarrow{D_\mu}\right]\psi(x)\,, \\
O_5(x) &=& m^2\psibar(x)\psi(x)\,, \\
O_6(x) &=& \lambdabar(x)\lambda(x)\,, \\
O_7(x) &=& m \left(\lambdabar(x)\psi(x) + \psibar(x)\lambda(x)\right)\,, \\
O_8(x) &=& \lambdabar(x)\gamma_\mu D_\mu\psi(x) - 
\psibar(x)\gamma_\mu\overleftarrow{D_\mu}\lambda(x)\,, \\
O_9(x) &=& \left.\partial_t\left(\chibar(x,t)\chi(x,t)\right)\right|_{t=0}\,,
\label{eq:list_fields}
\eea 
where the first $5$, $O_1, \ldots ,O_5$, are the standard terms 
from the unflowed theory~\cite{Luscher:1996sc},
and the additional $4$, $O_6, \ldots ,O_9$, are the new contributions 
stemming from the gradient flow equation. 
For on-shell O($a$) improvement one can use the field equations for 
$\psi$, (and $\psibar$), $\chi$, (and $\chibar$),
while the field equations for $\lambda$, (and $\lambdabar$) are equivalent to impose
the gradient flow equation. 
The total number of conditions is $4$, leaving $5$ total independent fields.
From the first $5$ fields, $O_1-O_5$,
I make the standard choice~\cite{Luscher:1996sc} to select $O_1$, $O_3$ and $O_5$.
In Ref.~\cite{Luscher:2013cpa}, for the additional fields $O_6-O_9$, 
the choice is to select $O_6$ and $O_7$.
The field $O_6 = \lambdabar\lambda$ is multiplied by the improvement coefficient 
$\cfl$, while $O_7$ is responsible for the mass dependent cutoff effects
removed by the improvement coefficients $b_\chi$.\footnote{With
more than one flavor there is an additional $D=5$ field 
responsible to the term proportional to $\overline{b}_\chi$.}
In this study I select instead $O_7$ and $O_8$.
Our choice is dictated by the following observation. 
If I write explicitly the terms of the summation over 
$\epsilon$ of Eq.~\eqref{eq:SFfl_latt}
\bea 
S_{\rm{F,fl}} &=& a^4 \sum_x\left[\lambdabar(x)\chi(x,\epsilon) - \lambdabar(x) \chi(x,t=0) - 
\epsilon \lambdabar(x) \nabla^2 \chi(x,t=0) + \right.\nonumber  \\
&+& \left. \chibar(x,\epsilon) \lambda(x) -\chibar(x,t=0) \lambda(x) 
- \epsilon \psibar(x) \nabla^2 \lambda(x)\right] + \cdots\,.
\label{eq:gf_ferm_epsilon}
\eea
the second and the fifth terms contain the fermion fields 
defined by the initial conditions of the 
gradient flow equations.
The lattice version of the fields $O_7$ and $O_8$ can then be 
included in the lattice action 
modifying the initial conditions.
If I now modify the initial conditions 
\bea 
\chi(x,t)|_{t=0} = (1+\frac{a}{2}c_1\gamma_\mu D_\mu + \frac{a}{2} c_2 m)\psi(x)\,, \\
\chibar(x,t)|_{t=0} = \psibar(x)(1-\frac{a}{2}c_1\gamma_\mu \overleftarrow{D}_\mu + \frac{a}{2} c_2 m) \nonumber \,.
\label{eq:new_bc}
\eea 
the second and fifth terms in Eq.~\eqref{eq:gf_ferm_epsilon} 
change as follows
\bea 
\lambdabar(x) \chi(x,t=0) \rightarrow \lambdabar(x)(1+\frac{a}{2}c_1\gamma_\mu D_\mu + \frac{a}{2} c_2 m)\psi(x)\,, \\
\chibar(x,t=0) \lambda(x) \rightarrow \psibar(x)(1-\frac{a}{2}c_1\gamma_\mu \overleftarrow{D}_\mu + \frac{a}{2} c_2 m)\lambda(x)\,.
\eea 
The modified form of the action $S_{\rm{F,fl}}$ now contains automatically 
the fields $O_7$ and $O_8$.

The conclusion is that the O($a$) improvement of the theory can be obtained modifying the initial
conditions at finite lattice spacing. With this formulation one does not need to determine 
the coefficient $\cfl$ and one does not need to compute additional correlation functions
with the space-time insertion of the term $\cfl \lambdabar \lambda$.

\section{Tree-level analysis}
\label{sec:tl}

To study cutoff effects I first 
consider standard Wilson fermions at tree-level of perturbation 
theory. In the next Sec.~\ref{ssec:tl_gf} I extend this analysis to include
flowed fermion fields.

In momentum space the standard Wilson fermion tree-level propagator is given by 
\be 
\WS_{\text{W}}(p) = \frac{- i \pall{\newsl{p}} + M(p)}{\pall{p}^2 + M(p)^2}\,,
\ee
where $M(p) = m + \frac{1}{2}a \pcap ^2$, 
$\ppall_\mu = \frac{1}{a}\sin(a p_\mu)$ and 
$\pcap_\mu = \frac{2}{a}\sin(\frac{a p_\mu}{2})$.
The only step needed to renormalize the quark propagator
is a redefinition of the quark mass.
Using the pole mass definition, $m \rightarrow m(1+\frac{1}{2}am)$,
it is equivalent to include the field $\mcO_5$ in the lattice theory.
At leading order in the lattice spacing $a$ the quark propagator now reads 
(see for example Ref.~\cite{Heatlie:1990kg,Capitani:2000xi})
\be 
S(x,y) \rightarrow \int \frac{d^4p}{(2 \pi)^4}\e^{i p (x-y)}
\frac{-i \newsl{p}+m}{p^2 + m^2}(1-am) + \frac{1}{2}a \delta^{(4)}(x-y) + O(a^2)\,.
\label{eq:propexp}
\ee
The last constant term is a contact term term 
proportional to $\delta^{(4)}(x-y)$,
while the residual O($am$) contribution can be removed improving the observable, i.e. 
the fermion fields in this case.
Improving the fermion fields
\be 
\begin{cases}
    \psi_{\text{I}}(x) = \psi(x)\left(1+\frac{a}{2}b_\psi m\right)\\
    \psibar_{\text{I}}(x) = \psibar(x)\left(1+\frac{a}{2}b_\psi m\right)\,,
\end{cases}
\ee
with the tree-level value $b_\psi^{(0)} = 1$,
the improved propagator reads
\be 
\left\langle \psi_{\text{I}}(x)\psibar_{\text{I}}(y)\right\rangle = S_{\text{I}}(x-y) = 
\int \frac{d^4p}{\left(2\pi\right)^4}\frac{- i \newsl{p} + m }{p^2 + m^2} + 
\frac{1}{2}a\delta^{(4)}(x-y) +O(a^2)\,.
\ee 
This result confirms the expectation of the Symanzik program. I have improved 
the theory and the observable and I obtain an O($a$) improved result, excluding 
contact terms.\footnote{To remove also the contact term one can modify the fermion 
field, 
$\psi(x) \rightarrow \left(1 + \frac{a}{4} c_q \left(\newsl{D} + m\right)\right)\psi(x)$,
with tree-level value $c_q^{(0)} = -1$, 
as discussed in Ref.~\cite{Heatlie:1990kg,Martinelli:2001ak}.}

\subsection{Tree-level analysis of the flowed fermion propagator}
\label{ssec:tl_gf}

At finite lattice spacing 
the flowed fermion propagator is computed solving the 
discretized version of the gradient flow equation~\eqref{eq:gf_ferm_epsilon}
\be 
\WS_{\text{W}}(p,t,s) = 
\e^{-\phat^2(t+s)}\frac{- i \pall{\newsl{p}} + M(p)}{\pall{p}^2 + M(p)^2}\,.
\label{eq:propwflow}
\ee 
Expanding in powers of $a$ and rescaling the quark mass, $m \rightarrow m(1+1/2 am)$, 
one obtains
\be 
\WS_{\text{W}}(p,t,s) = \e^{-p^2(t+s)}\left(1+\frac{a^2 p^2}{12}(t+s)\right)
\frac{- i \newsl{p} + m }{p^2 + m^2}\left(1-am\right) + 
\frac{1}{2}a\e^{-p^2(t+s)}\left(1+\frac{a^2 p^2}{12}(t+s)\right) + \cdots\,,
\label{eq:propwflowexp}
\ee 
where, beside the O($a$), the equation shows also  
the O($a^2$) resulting from the expansion 
of the $\nabla^2$ term. Modifying the gradient flow differential operator 
as discussed earlier, $\nabla_\mu^*\nabla_\mu \rightarrow 
\nabla_\mu^*\nabla_\mu\left(1-\frac{a^2}{12} \nabla_\mu^* \nabla_\mu \right)$
subtracts those particular O($a^2$) effects. 
Only numerical experiments can test the effectiveness 
to use the improved laplacian operator and first numerical 
tests have been shown in Ref.~\cite{Battelli:2022kbe}.

I now drop all the O($a^2$) terms and continue the analysis retaining
from Eq.~\eqref{eq:propwflowexp} only the O($a$) terms.
Following Ref.~\cite{Luscher:2013cpa} to improve the 
observable I first need to improve the fermion fields as follows
\be 
\begin{cases}
    \chi_{\text{I}}(x,t) = \left(1+\frac{a}{2}b_\chi m\right)\chi(x,t)\\
    \chibar_{\text{I}}(x,t) = \chibar(x,t)\left(1+\frac{a}{2}b_\chi m\right)\,,
\end{cases}
\ee
with a tree-level value $b_\chi^{(0)} = 1$. The propagator now is 
\be 
\WS_{\text{I}}(p,t,s) = \e^{-p^2(t+s)}
\frac{- i \newsl{p} + m }{p^2 + m^2} + 
\frac{1}{2}a\e^{-p^2(t+s)} + O(a^2)\,.
\label{eq:propiflowexp}
\ee 
The propagator is still affected by O($a$) cutoff effects which are the remnant of the 
contact term in Eq.~\eqref{eq:propexp}. The gradient flow regulates the contact
term generating a new O($a$) term parametrized, in the Symanzik effective theory, 
by a new $D=5$ field. Following Ref.~\cite{Luscher:2013cpa} the additional 
O($a$) cutoff effects are removed tuning the coefficient 
of $O_6 = \lambdabar \lambda$,
denoted as $\cfl$. In practice the term $\cfl O_6$ is inserted 
in the correlation functions with tree-level value $\cfl^{(0)} = 1/2$.

I now show that the same cancellation takes place 
modifying the initial boundary conditions as discussed in Sec.~\ref{sec:gf}
(see Eq.~\eqref{eq:new_bc}).
With the new boundary conditions~\eqref{eq:new_bc} 
the lattice flowed fermion propagator is 
\be 
\WS(p,t,s) = 
\e^{-\phat^2(t+s)}\left(1+\frac{a}{2}c_1^{(0)}i\pall{\newsl{p}} + \frac{a}{2}c_2^{(0)} m\right)
\left(i \pall{\newsl{p}} + M(p)\right)^{-1}\left(1+\frac{a}{2}c_1^{(0)}i\pall{\newsl{p}} + \frac{a}{2}c_2^{(0)} m\right)\,.
\label{eq:newpropwflow}
\ee 
After rescaling the quark mass, the remaining O($a$) effects in the propagator
are removed tuning the tree-level values of the improvement
coefficients to $c_1^{(0)} = -1/2$ and $c_2^{(0)} = 1/2$.
It is maybe convenient to rewrite the initial conditions as 
\bea 
\chi(x,t)|_{t=0} = \left(1+\frac{a}{2}c_\chi\left(\gamma_\mu D_\mu + m\right) + 
\frac{a}{2} c_m m\right)\psi(x)\,, \label{eq:new_bc2} \\
\chibar(x,t)|_{t=0} = \psibar(x)\left(1-\frac{a}{2}c_\chi\left(\gamma_\mu \overleftarrow{D}_\mu + m \right)
+ \frac{a}{2} c_m m\right) \nonumber \,,
\eea 
where $c_\chi = c_1$ and $c_m = c_2 - c_1$, 
with tree-level values $c_\chi^{(0)} = -1/2$ 
and $c_m^{(0)} = 1$. 
It is possible to show that the improvement coefficients 
$c_\chi$ and $c_m$ are related to $\cfl$ and $b_\chi$.
The term proportional to $c_\chi$ 
can be implemented numerically using any lattice form of the Dirac operator,
while the term proportional to $c_m$ can be either included in the 
initial conditions as in Eq.~\eqref{eq:new_bc2} or as a multiplicative 
factor in flowed correlators as done in Ref.~\cite{Luscher:2013cpa} with $b_\chi$.
A form of the initial conditions that avoids including the quark mass is 
\bea 
\chi(x,t)|_{t=0} = \left(1+\frac{a}{2}c_\chi \gamma_\mu D_\mu \right)\psi(x)\,, 
\label{eq:new_bc3} \\
\chibar(x,t)|_{t=0} = \psibar(x)\left(1-\frac{a}{2}c_\chi \gamma_\mu \overleftarrow{D}_\mu 
\right) \nonumber \,,
\eea 
where also in this case the term proportional to the mass is 
added to the correlation functions but with a different improvement coefficient 
than $b_\chi$.
Different choices of the lattice Dirac operator
in the initial conditions generate different higher order cutoff effects 
and only numerical studies can provide an indication on 
which choice is better in terms of O($a^2$) effects.

\section{Final remarks}
\label{sec:final}

The gradient flow for Wilson-type fermions simplifies the 
process of renormalization and improvement of flowed 
correlators~\cite{Luscher:2013cpa}. Matrix elements of flowed operators 
renormalize multiplicatively and can be improved at the classical level,
provided the lattice QCD action at the boundary $t=0$ is O($a$) improved. 
The price to pay is some additional 
O($a$) boundary, $t=0$, terms which are related to the use of flowed fermion fields.
I have shown that these O($a$) effects are a remnant of the 
O($a$) proportional to contact terms in the unflowed theory.
I have also shown that using the modified gradient flow equation 
\bea 
&& \partial_t \chi(x,t) = \nabla_\mu^*\nabla_\mu\left(1-
\frac{a^2}{12} \nabla_\mu^* \nabla_\mu \right) \chi(x,t)\,, \label{eq:new_gf} \\
&& \chi(x,t)|_{t=0} = \left(1+\frac{a}{2}c_\chi \gamma_\mu D_\mu\right)\psi(x) \,, \nonumber
\eea 
and the corresponding for $\chibar$, flowed observables are O($a$) improved,
provided the lattice QCD action is also non-perturbatively O($a$) improved.
The modified lattice version of the Laplacian in Eq.~\eqref{eq:new_gf} is O($a^2$)
improved~\cite{Battelli:2022kbe}. Only numerical tests can indicate whether 
the use of the improved Laplacian and the specific choice of the 
lattice covariant derivatives is useful to decrease discretization errors.
The remaining O($am$) terms are removed
multiplying flowed correlators with the proper rescaling factor for
fermion fields as discussed in Ref.~\cite{Luscher:2013cpa}.
With the GF equations~\eqref{eq:new_gf}
it is not necessary to determine additional correlation functions
to have an O($a$) improved lattice theory~\cite{Luscher:2013cpa}.
It would be interesting to test 
if chiral Ward identities can be used to estimate $c_\chi$ as 
done for $\cfl$.

\section*{Acknowledgments}
\label{sec:ack}
I acknowledge funding support under the National Science Foundation 
grant PHY-2209185.

\bibliographystyle{utphysmod}
\bibliography{gf_impr.bib}



\end{document}